\documentclass[12pt,preprint]{aastex}
 \slugcomment{2017, ApJL, 835, L38}
 \shorttitle{The Spectral Energy Distribution of Fermi Blazars}
 \shortauthors{Fan et al.}

\begin{document}
\title{Intrinsic Correlations for Flaring Blazars Detected by Fermi}

\author{J. H. Fan$^{1,2}~${\thanks{email:fjh@gzhu.edu.cn}}, J. H. Yang$^3${\thanks{email:yjianghe@163.com}}, H. B. Xiao$^{1,2}$, C. Lin$^{1,2}$, D. Constantin$^{1,2}$, G. Y. Luo$^{1,2}$, Z.Y. Pei$^{1,2}$, J. M. Hao$^{1,2}$, Y. W. Mao$^{1,2}$}
\affil{
1. Center for Astrophysics, Guangzhou University, Guangzhou 510006, China\\
2. Astronomy Science and Technology Research Laboratory of
   Department of Education of Guangdong Province, Guangzhou 510006,
   China\\
3.Department of Physics and Electronics Science, Hunan University
of Arts and Science, Changde 415000, China
}

\begin{abstract}
Blazars are an extreme subclass of active galactic nuclei. Their rapid variability, luminous brightness, superluminal motion, and high and variable polarization are probably due to  a beaming effect. However,  this beaming factor (or Doppler factor) is  very difficult to measure.
Currently, a good way to estimate  it is to use the time scale of their radio flares.
 In this $Letter$, we use  multiwavelength data and Doppler factors  reported in the literatures  for a sample of 86 flaring blazars detected by Fermi to compute  their intrinsic multiwavelength data and intrinsic spectral energy distributions, and investigate the correlations among observed and intrinsic data.
  Quite interestingly, intrinsic data show a positive correlation between luminosity and peak frequency, in contrast with the behavior of observed data, and a tighter correlation between  $\gamma$-ray luminosity and the lower energy ones.
 For flaring blazars detected by Fermi, we conclude that
 (1) Observed emissions are strongly beamed;
 (2) The anti-correlation between luminosity and peak frequency from the observed data is an apparent result,
  the correlation between intrinsic data being positive; and
 (3) Intrinsic $\gamma$-ray luminosity is strongly correlated with
 other intrinsic luminosities.

\end{abstract}



\section{Introduction}

 Blazars show extreme observational properties
 (Acero et al. 2015;
 Ackermann et al. 2015;
 Fan et al. 2016a,b; and reference therein). They are divided into two main subclasses,
   BL Lacertae objects (BL Lacs) and
 flat spectrum radio quasars (FSRQs), following to the behavior of their emission lines and further subdivided according to their synchrotron peak frequency
   (
   Padovani \& Giommi 1995;
   Nieppola et al. 2006;
   Abdo et al. 2010;
   Fan et al. 2016a,c
   ).
 Recently, Massaro et al. (2015) published the largest blazar sample (the BZCAT 5.0 (http://www.asdc.asi.it/bzcat/).

Blazars were detected by EGRET
(
 Hartman et al. 1999
 ) and are the main discovery of Fermi/LAT mission
 (
 Abdo et al. 2009;
 Nolan et al. 2012;
 Acero et al. 2015;
 Ackermann et al. 2015
 ).
 The correlation between  $\gamma$-rays and radio bands suggests a strong beaming effect in $\gamma$-rays
    \citep{Dondi1995, Fan1998, Huang1999,Cheng2000,Fan2016a}, and the beaming factors (or Doppler factors) are estimated  for some $\gamma$-ray loud
   blazars based on their rapid $\gamma$-ray variability time scale, X-ray and $\gamma$-ray emissions (
     Mattox et al.  1993,
     von Montigny et al. 1995,
     Cheng, et al. 1999;
     Fan et al. 1999; 2009; 2013; 2014;
     Fan 2005
     ).
 In addition,  a useful method to estimate Doppler factors may be found in a work  \citep{Lahteenimaki1999}, followed by many other works
\citep{Fan2009, Hovatta2009, Lister2009a, Savolainen2010}.
 Doppler factor is also obtained by a synchrotron self-Compton mechanism
\citep{Ghisellini1993}.

 The strong beaming effect could be the reason of the correlations among different energy bands and among the observed parameters. For this reason, we investigate whether such correlations hold even for the intrinsic emission.
 We adopt
      $f^{in}\, = f^{ob}/\delta^{p},$
  where
   $f^{in}$ is the intrinsic (or de-beamed) flux density,
   $f^{ob}$ is the observed flux density,
   $\delta$ is a Doppler factor (or boosting factor),
   $p = 2 + \alpha$ for a continuous jet (or
   $p = 3 + \alpha$ for a spherical  jet), and
   $\alpha$ is a spectral index ($f_{\nu} \propto \nu^{-\alpha}$).

 In this $Letter$ we investigate the relationship between the observed $\gamma$-ray luminosity and the peak frequency and  those among their intrinsic values for a sample of Fermi blazars (Fan et al. 2016a) with available Doppler factors. The sample properties are presented in section 2, the results in section 3 while in section 4 we give our conclusions.

\section{Sample and Results}


In this $Letter$, a sample of  86 flaring blazars detected by Fermi (55 FSRQs and 31 BLs) with available Doppler factor \citep{Lahteenimaki1999, Fan2009, Hovatta2009, Lister2009a, Savolainen2010} is compiled. Their synchrotron peak frequency,
 multi-wavelength monochromatic luminosity, and peak luminosity are from our recent paper (Fan et al. 2016a). The radio data is at 1.4 GHz, optical data at R band, and the X-ray data  at 1 KeV.
But the X-ray flux density is only available for 82 sources, the corresponding multiwavelength luminosities are shown in Table \ref{sample}.

\begin{deluxetable}{lcccccccccccc}
\tabletypesize{\scriptsize}
 \rotate
  \tablecaption{Sample of Fermi blazars with Doppler Factors}
  \tablewidth{0pt}
 \tablehead{
  \colhead{ 3FGL name }&
  \colhead{ Other name } &
  \colhead{ redshift } &
  \colhead{ Class } &
  \colhead{ log $\nu_p$ } &
  \colhead{ log$\rm L_p$  } &
  \colhead{ log$\rm L_b$  }&
  \colhead{ log$\rm L_R$  }&
  \colhead{ log$\rm L_O$  }&
  \colhead{ log$\rm L_X$  }&
  \colhead{ log$\rm L_{\gamma}$  }&
   \colhead{ $\delta_R$  }&
   \colhead{ Ref  }
    }
 \startdata

3FGL J0050.6-0929	&	PKS 0048-09	&	0.635 	&	IBL	&	14.60 	&	45.99 	&	46.46 	&	43.05 	&	45.75 	&	45.30 	&	46.04 	&	9.6	&	 H09	\\
3FGL J0108.7+0134	&	4C +01.02	&	2.099 	&	IF	&	13.53 	&	46.47 	&	47.00 	&	44.57 	&	46.29 	&	45.47 	&	47.66 	&	18.2	 &	S10	\\
3FGL J0112.1+2245	&	S2 0109+22	&	0.265 	&	IBL	&	14.39 	&	45.40 	&	45.73 	&	41.94 	&	45.36 	&	44.53 	&	45.41 	&	9.1	&	 S10	\\
3FGL J0137.0+4752	&	OC 457	&	0.859 	&	LF	&	12.69 	&	46.14 	&	46.41 	&	43.47 	&	45.50 	&	45.05 	&	46.55 	&	20.5	&	 S10	\\
3FGL J0151.6+2205	&	PKS 0149+21	&	1.320 	&	LF	&	13.14 	&	46.09 	&	46.48 	&	43.80 	&	45.59 	&		&	46.23 	&	4.72	&	 LV99	\\
..... & .....&..... & .....&..... & .....&..... & .....&..... & .....&..... & .....&..... \\
..... & .....&..... & .....&..... & .....&..... & .....&..... & .....&..... & .....&..... \\
..... & .....&..... & .....&..... & .....&..... & .....&..... & .....&..... & .....&..... \\
..... & .....&..... & .....&..... & .....&..... & .....&..... & .....&..... & .....&..... \\
\enddata

Note to the Table:
Col. (1) gives the 3FGL name;
Col. (2) Other name;
Col. (3) redshift from NED database at IPAC;
Col. (4)  the SED  classification  by our method (Fan et al. 2016a)
  B stands for BL Lac,
  F for  FSRQ,
  BCU for unidentified source;
Col. (5)   peak frequency, $log \nu_p$ (Hz);
Col. (6)  peak luminosity, $\rm log L_p$ (erg/s);
Col. (7)  bolometric luminosity, $\rm log L_{bol}$ (erg/s);
Col. (8)  radio luminosity, $\rm log L_R$ (erg/s);
Col. (9)  optical luminosity, $\rm log L_O$ (erg/s);
Col. (10)  X-ray luminosity, $\rm log L_X$ (erg/s);
Col. (11)  $\gamma$-ray luminosity, $\rm log L_{\gamma}$ (erg/s);
Col. (12)  Doppler factor, $\rm \delta_R$; and
Col. (13)  reference for Doppler factor.
          F09: Fan et al. (2009);
          H09: Hovatta et al. (2009);
          LV99: L\"ahteenim\"aki \& Valtaoja (1999);
           S10: Savolainen et al. (2010)
\label{sample}
\end{deluxetable}

\begin{figure}
    \centering
    \resizebox{\hsize}{!}{\includegraphics*{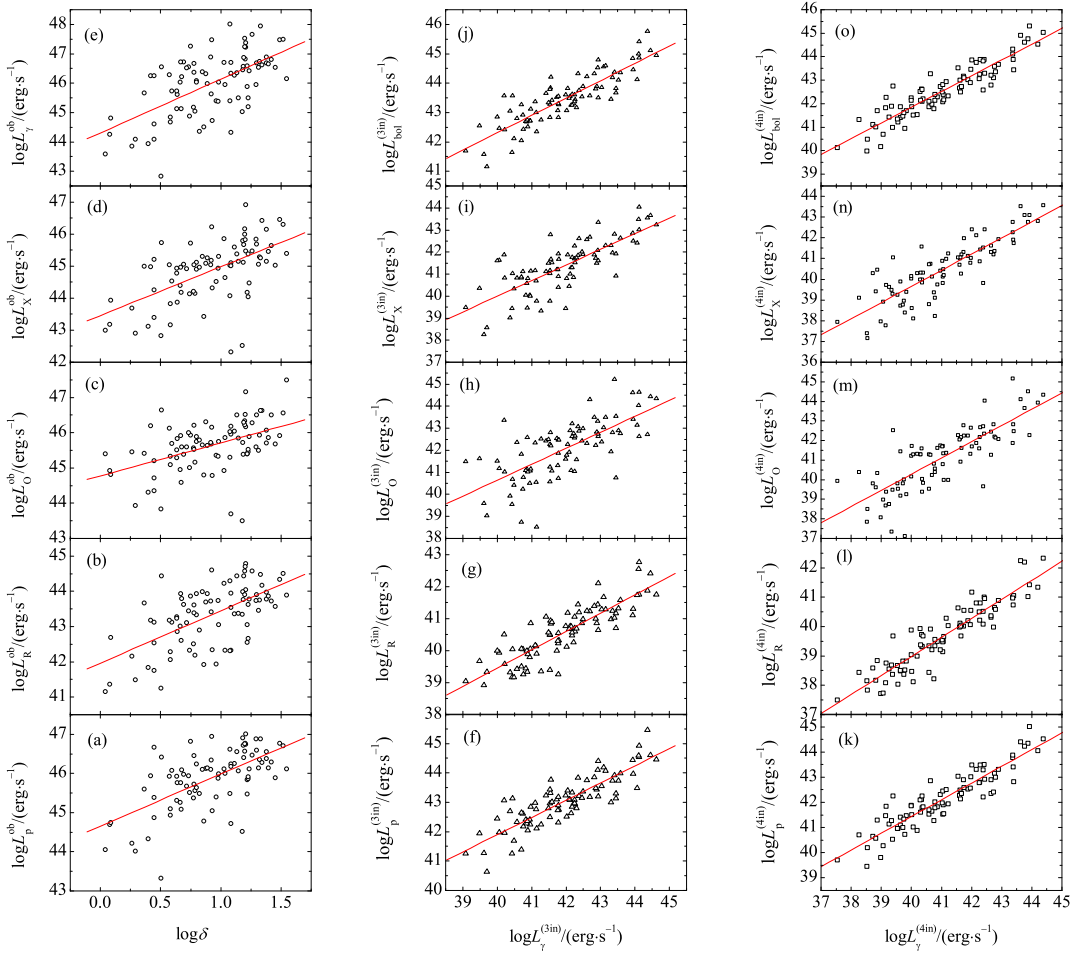}}
    \caption{Left panel: The correlation between   luminosity ($\rm log {\nu L_{\nu}}$) and peak frequency ( $\rm log \nu_p$). From the top to the bottom is for $\gamma$-ray, X-ray, optical, radio, and peak luminosity;
       Middle panel: Correlation for intrinsic values between $\gamma$-ray and monochromatic luminosity for the case of $p = 2 + \alpha$
       (from the top to the bottom is for bolometric, X-ray, optical, radio, and peak luminosity), and
       Right panel: Correlation for intrinsic values between $\gamma$-ray and monochromatic luminosity for $p = 3 + \alpha$. }
     \label{Fan-ApJL-2016-fig1}
\end{figure}

\begin{figure}
    \centering
    \resizebox{\hsize}{!}{\includegraphics*{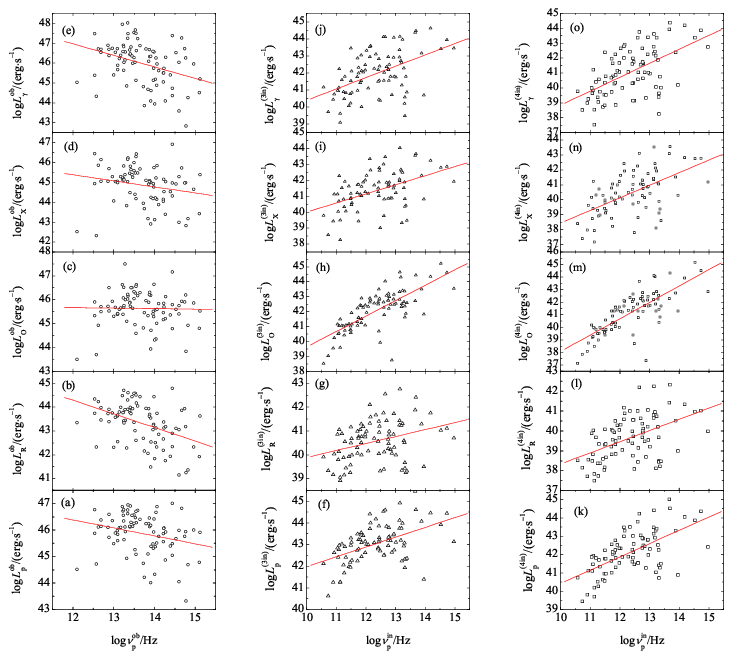}}
    \caption{The correlation between   luminosity ($\rm log {L_{\nu}}$) and peak frequency ( $\rm log \nu_p$).
     From the upper one to the bottom one is for $\gamma$-ray, X-ray, optical, radio, and peak luminosities.
       Left panel: observed values,
       Middle panel: intrinsic values for the case of $p = 2 + \alpha$,
       Right panel:  intrinsic values for the case of $p = 3 + \alpha$. }
    \label{Fan-ApJL-2016-fig2}
\end{figure}

 For the sample, we investigate the relationship between luminosity and Doppler factor and show the results in Table \ref{result} and Fig. \ref{Fan-ApJL-2016-fig1}.
For the relationship between   the peak frequency, $ log \nu$, and  luminosity, the corresponding  results are listed in Table \ref{result} and shown in Fig. \ref{Fan-ApJL-2016-fig2}.

When we consider the intrinsic ( or de-beamed ) monochromatic luminosity, we have
 $\rm log (\nu L_{\nu}^{in}) = log (\nu L_{\nu})$ -  $(p + 1)\,log \delta. $
  Here we obtain the spectral index, $\alpha$ by fitting the data ($f_{\nu}\, \propto \, \nu^{-\alpha}$)  at the corresponding bands. We  calculate radio spectral indexes ($\alpha_R$) for all the 86 sources,  optical indexes ($\alpha_O$) for 62 sources, and X-ray spectral indexes ($\alpha_X$) for only 5 sources. For the sources without fitting $\alpha_O$ and $\alpha_X$, we look for those parameters from the available literatures and get 70 $\alpha_X$'s and 4 $\alpha_O$'s. Then we have averaged spectral index, $<\alpha_O>$ = 1.155 for 28 BL Lacs and $<\alpha_O>$ = 0.805 for 38 FSRQs; $<\alpha_X>$ = 1.021 for 25 BL Lacs and $<\alpha_X>$ = 0.738 for 50 FSRQs, they are used to replace unknown spectral indexes.

  For the intrinsic peak frequency,  peak flux, and integrated flux, we use the intrinsic multiwavelength data to calculate the SEDs as did in Fan et al. (2016a),
  $\rm log\, (\nu^{in} f^{in}_{\nu}) = -P_1(log \nu^{in}\, - \, P_2)^2\, + \, P_3$,
  here $f^{in} = f^{ob}/\delta^p$, $\nu^{in}=\nu/(\delta/(1+z))$,
  $\rm P_1$ is a curvature parameter,
  $\rm P_2$ is the intrinsic peak frequency, and
  $\rm P_3$ is the intrinsic integrated flux, from which we calculate the intrinsic bolometric luminosity.

The correlation between intrinsic $\gamma$-luminosity and  monochromatic ( peak and bolometric ) luminosity are shown in Fig.  \ref{Fan-ApJL-2016-fig1} and those between the intrinsic luminosity and peak frequency are  listed in Table \ref{result} and shown in Fig.  \ref{Fan-ApJL-2016-fig2}.

\begin{deluxetable}{lccccc}
\tabletypesize{\scriptsize}
  \tablecaption{Linear Correlation Fitting Results, $y = a\, + \, b x$ }
  \tablewidth{0pt}
 \tablehead{
  \colhead{ $y\, \sim \, x$}&
  \colhead{ $a\,\sim \, \Delta a$ } &
  \colhead{ $b\,\sim \, \Delta b$ } &
  \colhead{ N } &
  \colhead{ $r$ } &
  \colhead{ $p$  }
    }
 \startdata
log $L_p^{ob}\,\, \sim \, log \delta$ 	&	44.656 	$\pm$	0.178 	&	1.330 	$\pm$	0.181 	&	86	&	0.625 	&	1.22$\times10^{-10}$	\\
log $L_R^{ob}\,\, \sim \, log \delta$ 	&	41.963 	$\pm$	0.209 	&	1.484 	$\pm$	0.212 	&	86	&	0.606 	&	6.14$\times10^{-10}$	\\
log $L_o^{ob}\,\, \sim \, log \delta$ 	&	44.775 	$\pm$	0.190 	&	0.935 	$\pm$	0.193 	&	86	&	0.468 	&	5.62$\times10^{-6}$	\\
log $L_X^{ob}\,\, \sim \, log \delta$ 	&	43.456 	$\pm$	0.239 	&	1.521 	$\pm$	0.242 	&	82	&	0.575 	&	1.60$\times10^{-8}$	\\
log $L_{\gamma}^{ob}\,\, \sim \, log \delta$ 	&	44.304 	$\pm$	0.252 	&	1.832 	$\pm$	0.257 	&	86	&	0.614 	&	3.15$\times10^{-10}$	 \\
log $L_p^{3in}\,\, \sim \, log L_{\gamma}^{3in}$ 	&	18.472 	$\pm$	1.603 	&	0.586 	$\pm$	0.038 	&	86	&	0.858 	&	 4.49$\times10^{-26}$	\\
log $L_R^{3in}\,\, \sim \, log L_{\gamma}^{3in}$	&	16.675 	$\pm$	1.561 	&	0.570 	$\pm$	0.037 	&	86	&	0.858 	&	 4.86$\times10^{-26}$	\\
log $L_O^{3in}\,\, \sim \, log L_{\gamma}^{3in}$	&	11.650 	$\pm$	3.399 	&	0.725 	$\pm$	0.081 	&	86	&	0.699 	&	 7.36$\times10^{-14}$	\\
log $L_X^{3in}\,\, \sim \, log L_{\gamma}^{3in}$	&	11.591 	$\pm$	2.752 	&	0.710 	$\pm$	0.066 	&	82	&	0.771 	&	 2.43$\times10^{-17}$	\\
log $L_{bol}^{3in}\,\, \sim \, log L_{\gamma}^{3in}$	&	18.698 	$\pm$	1.572 	&	0.590 	$\pm$	0.037 	&	86	&	0.864 	&	 8.03$\times10^{-27}$	\\
log $L_p^{4in}\,\, \sim \, log L_{\gamma}^{4in}$ 	&	14.822 	$\pm$	1.314 	&	0.665 	$\pm$	0.032 	&	86	&	0.915 	&	 6.91$\times10^{-35}$	\\
log $L_R^{4in}\,\, \sim \, log L_{\gamma}^{4in}$	&	13.049 	$\pm$	1.261 	&	0.648 	$\pm$	0.031 	&	86	&	0.917 	&	 2.35$\times10^{-35}$	\\
log $L_O^{4in}\,\, \sim \, log L_{\gamma}^{4in}$	&	7.134 	$\pm$	2.806 	&	0.829 	$\pm$	0.068 	&	86	&	0.798 	&	 3.68$\times10^{-20}$	\\
log $L_X^{4in}\,\, \sim \, log L_{\gamma}^{4in}$	&	8.452 	$\pm$	2.242 	&	0.780 	$\pm$	0.055 	&	82	&	0.847 	&	 1.04$\times10^{-23}$	\\
log $L_{bol}^{4in}\,\, \sim \, log L_{\gamma}^{4in}$	&	15.052 	$\pm$	1.293 	&	0.670 	$\pm$	0.031 	&	86	&	0.918 	&	 1.39$\times10^{-35}$	\\
log $L_p^{ob}\,\, \sim \, log \nu_p^{ob}$	&	49.986 	$\pm$	1.678 	&	-0.300 	$\pm$	0.122 	&	86	&	-0.259 	&	1.6\%	\\
log $L_R^{ob}\,\, \sim \, log \nu_p^{ob}$	&	51.138 	$\pm$	1.808 	&	-0.571 	$\pm$	0.132 	&	86	&	-0.427 	&	4.15$\times10^{-5}$	\\
log $L_O^{ob}\,\, \sim \, log \nu_p^{ob}$	&	45.921 	$\pm$	1.633 	&	-0.021 	$\pm$	0.119 	&	86	&	-0.019 	&	86\%	\\
log $L_X^{ob}\,\, \sim \, log \nu_p^{ob}$	&	49.027 	$\pm$	2.197 	&	-0.304 	$\pm$	0.160 	&	82	&	-0.208 	&	6.1\%	\\
log $L_{\gamma}^{ob}\,\, \sim \, log \nu_p^{ob}$	&	53.918 	$\pm$	2.276 	&	-0.580 	$\pm$	0.166 	&	86	&	-0.356 	&	7.73$\times10^{-4}$	 \\
log $L_p^{3in}\,\, \sim \, log \nu_p^{3in}$	&	37.472 	$\pm$	1.081 	&	0.451 	$\pm$	0.087 	&	86	&	0.492 	&	1.53$\times10^{-6}$	\\
log $L_R^{3in}\,\, \sim \, log \nu_p^{3in}$	&	36.971 	$\pm$	1.141 	&	0.292 	$\pm$	0.092 	&	86	&	0.327 	&	2.13$\times10^{-3}$	\\
log $L_O^{3in}\,\, \sim \, log \nu_p^{3in}$	&	29.299 	$\pm$	1.268 	&	1.033 	$\pm$	0.102 	&	86	&	0.740 	&	3.71$\times10^{-16}$	\\
log $L_X^{3in}\,\, \sim \, log \nu_p^{3in}$	&	34.302 	$\pm$	1.496 	&	0.571 	$\pm$	0.121 	&	82	&	0.468 	&	9.31$\times10^{-6}$	\\
log $L_{\gamma}^{3in}\,\, \sim \, log \nu_p^{3in}$	&	33.734 	$\pm$	1.581 	&	0.666 	$\pm$	0.128 	&	86	&	0.495 	&	1.26$\times10^{-6}$	 \\
log $L_p^{4in}\,\, \sim \, log \nu_p^{4in}$	&	33.228 	$\pm$	1.285 	&	0.720 	$\pm$	0.104 	&	86	&	0.604 	&	7.26$\times10^{-10}$	\\
log $L_R^{4in}\,\, \sim \, log \nu_p^{4in}$	&	32.727 	$\pm$	1.372 	&	0.561 	$\pm$	0.111 	&	86	&	0.484 	&	2.33$\times10^{-6}$	\\
log $L_O^{4in}\,\, \sim \, log \nu_p^{4in}$	&	25.056 	$\pm$	1.485 	&	1.302 	$\pm$	0.120 	&	86	&	0.765 	&	1.07$\times10^{-17}$	\\
log $L_X^{4in}\,\, \sim \, log \nu_p^{4in}$	&	30.024 	$\pm$	1.704 	&	0.842 	$\pm$	0.137 	&	82	&	0.566 	&	3.08$\times10^{-8}$	\\
log $L_{\gamma}^{4in}\,\, \sim \, log \nu_p^{4in}$	&	29.490 	$\pm$	1.822 	&	0.935 	$\pm$	0.147 	&	86	&	0.570 	&	 9.94$\times10^{-10}$	\\
\enddata

Note to the Table:
Col. (1) gives relation. Here
        $3in$ stands for the case of $p\,=\,2\,+\, \alpha$, and
        $4in$  for  $p\,=\,3 + \alpha$;
Col. (2) intercept and the corresponding uncertainty;
Col. (3) slope and the corresponding uncertainty;
Col. (4) number of the sample;
Col. (5)  correlation coefficient; and
Col. (6)  chance probability.
\label{result}
\end{deluxetable}

\section{Discussions}

 AGNs  are  the most numerous population of detected sources in the Fermi mission,
  which  provide us with a good opportunity to study their emission mechanism  and beaming effects.

In 1998, Fossati, et al. calculated the spectral energy distribution for a sample of blazars ( RBLs, XBLs, and FSRQs), investigated the relationship between the radio luminosity (peak luminosity) and peak frequency, and found that there is a sequence for blazars with higher radio luminosity sources corresponding to lower frequency and lower luminosity sources to higher peak frequency. In 2008, Nieppola et al. investigated the correlation between peak frequency and peak luminosity from observed and intrinsic data.
 The anti-correlation from the observed data was not found in the intrinsic data.
 Giommi et al. (2012a,b) pointed out that the anti-correlation is due to a  selection effect. Their FSRQs sample was not showing any anti-correlation (Giommi et al. 2012a), while a simulation  produced an anti-correlation between the radio luminosity and synchrotron peak frequency (Giommi et al. 2012b).  Mao et al. (2016) calculated SEDs for a large sample of blazars from the Roma-BZCAT catalog, and found that the peak frequency increases when the radio ( and bolometric ) luminosity decreases.

In the present $Letter$, we compile a sample of Fermi blazars with available Doppler factors  and
 investigate the relationship between peak luminosity (monochromatic luminosity) and Doppler factor, and that  between luminosity and peak frequency. It is found that the monochromatic luminosity is closely correlated with Doppler factor, which confirms that the emissions in blazars are strongly beamed. For the correlation between luminosity and peak frequency,  a close anti-correlation is found for radio luminosity, which is consistent with the results for the observed radio luminosity  in Fossati et al. (1998) and Mao et al. (2016). Our result for the observed peak luminosity and peak frequency is consistent with that in Fossati et al. (1998) and Nieppola et al. (2008). It is also found that an anti-correlation exists for the  $\gamma$-ray band ($r = -0.356$ and $p = 7.73 \times 10^{-4}$). But there is no  correlation for  optical band ($p = 86\%$), the reason is perhaps that some host galaxies in BL Lacs contribute optical emission and that the accretion disk in FSRQs contributes optical emission, which dilute the anti-correlation.
 When intrinsic data are considered, we have positive correlations between monochromatic (and peak) luminosity and peak frequency. Our results confirm that found in Nieppola et al. (2008) for  peak luminosity and peak frequency, and show a positive correlation for the $\gamma$-ray luminosity and peak frequency as shown in  Fig. \ref{Fan-ApJL-2016-fig1} and Table \ref{result}.
From Table \ref{result}, it can be clearly seen that the  correlation coefficients for the correlations of $\gamma$-ray against the lower monochromatic luminosity (radio, optical, X-ray) and the peak luminosity in the case of  $p = 3 + \alpha$ are greater than those for the corresponding correlations in the case of  $p = 2 + \alpha$, and the chance probability for the correlation in the case  $p = 3 + \alpha$ is much lower than that in the case of  $p = 2 + \alpha$.  When we consider the correlation between intrinsic luminosity and peak frequency, similar results are obtained. So,
   it is concluded that correlations for the intrinsic data in the case of  $p = 3 + \alpha$ are closer than those in the case of $p = 2 + \alpha$ suggesting that our discussion favors a spherical jet.

For the above discussions, we use the averaged values to replace the unknown optical and X-ray spectral indexes.  These replacements may cause some errors in correlation analysis. Thus, we re-investigate the correlations only  for the sources with available spectral indexes, and it is found that the fitting slopes, the correlation coefficient and the chance probability introduce only little change.

 Analyzing the behavior of the luminosity versus the peak frequency, we find anti-correlations in the observer frame and  positive ones in intrinsic data, suggesting that the anti-correlation may arise from a beaming effect or, as discussed in the works by Giommi et al. (2012a,b), from a selection effect.  In fact the sample is relatively small, as we considered here only sources which underwent flaring episodes, needed to estimate the Doppler factors, making up thus a sample of less than 10\% of Fermi blazars.
 In our previous paper (Fan et al. 2016a), we calculated spectral energy distributions (SEDs) for 1392 Fermi blazars,
 out of which 999 have redshifts and only 86 have available Doppler factors, we calculate
 their average monochromatic luminosity at radio, optical, X-ray, and $\gamma$-ray bands, and have
$<{\rm log\, L_R\, (erg/s)}> = 42.28\pm1.26$,
$<{\rm log\, L_O\, (erg/s)}> = 45.24\pm 0.84$,
$<{\rm log\, L_X\, (erg/s)}> = 44.60\pm0.99$, and
$<{\rm log\, L_{\gamma}\, (erg/s)}> = 45.27\pm1.31$ for the 913 sources without Doppler factors and
$<{\rm log\, L_R\, (erg/s)}> = 43.32\pm0.87$,
$<{\rm log\, L_O\, (erg/s)}> = 45.63\pm 0.72$,
$<{\rm log\, L_X\, (erg/s)}> = 44.86\pm0.96$, and
$<{\rm log\, L_{\gamma}\, (erg/s)}> = 45.98\pm1.07$
for  the 86 sources with available Doppler factors. When a K-S test is adopted to the luminosity distributions of the sources with/without known Doppler factors, it is found that the chance probability for the two luminosity distributions to be from the same distribution is
$p\, =\,4.11\times10^{-9}$, $1.60\times10^{-6}$, $1.11\times10^{-3}$, and $1.93\times10^{-4}$ for their radio, optical, X-ray and $\gamma$-ray bands respectively. It indicates that the sources in our sample is obviously brighter than those without Doppler factors. So, for the sources
 with flaring events, their  anti-correlation between observed luminosity and the peak frequency is an apparent result, their intrinsic correlation is positive. Since our  Fermi blazar sample with known Doppler factor is small, we will try to compute more Doppler factors for the Fermi blazars,
 and re-do the analysis in the future.

In our previous work, we found that $\gamma$-ray luminosity is closely correlated with  other monochromatic luminosity (X-ray, optical, and radio bands), bolometric and peak luminosity for a sample of 1392 Fermi blazars. When the redshift effect is removed, the correlation still exists between $\gamma$-ray and optical, radio, peak, and bolometric luminosities (Fan et al. 2016a). When the intrinsic luminosities are used for the correlation analysis, it is found that there are strong correlations between $\gamma$-ray luminosity and other monochromatic and peak luminosities. As listed in Table \ref{result}, we can see that strong correlations exist between $\gamma$-ray luminosity and peak, radio, and bolometric luminosities for the case of $p = 3 + \alpha$, with  $r \sim 0.92$ and  $p \leq 6.91\times10^{-35}$.  We conclude by noting that strong correlations are consistent with the fact that GeV $\gamma$-rays are from SSC process in blazars.

\section{Conclusions}

In this $Letter$, we consider a sample of 86  flaring blazars, detected by Fermi, with available Doppler factors, calculate their intrinsic SEDs, and investigate some correlations.   Our conclusions are as follows:
1) The anti-correlation between luminosity and peak frequency for  the flaring sources
 with known Doppler factors  in the observer's frame maybe caused by a beaming effect;
 2) There is a positive correlation between intrinsic  monochromatic ($\gamma$-ray, X-ray, optical, and radio band) luminosity and peak frequency;
3) There are strong correlations between intrinsic $\gamma$-ray and other luminosity  suggesting that the GeV $\gamma$-rays are mainly from SSC;
4) Our analysis favors a spherical jet in the Fermi blazars.

\begin{acknowledgements}

  The authors thank the anonymous referee for the suggestions and comments.
 This work is supported by the National Natural Science Foundation of China (U1531245, U1431112, 11203007, 11403006, 10633010, 11173009, 11503004), and the Innovation Foundation of Guangzhou University (IFGZ), Guangdong Innovation Group for Astrophysics(2014KCXTD014),
 Guangdong Province Universities and Colleges Pearl River Scholar
 Funded Scheme(GDUPS)(2009), Yangcheng Scholar Funded
 Scheme(10A027S), and supported for Astrophysics  Key Subjects of Guangdong Province and Guangzhou City. This paper is dedicated to  Prof. Guangzhong Xie (G.Z. Xie) who passed away on 2016 August 19 in Kunming, China.

\end{acknowledgements}




\begin{deluxetable}{lcccccccccccc}
\tabletypesize{\scriptsize}
 \rotate
  \tablecaption{Sample of Fermi blazars with Doppler Factors}
  \tablewidth{0pt}
 \tablehead{
  \colhead{ 3FGL name }&
  \colhead{ Other name } &
  \colhead{ redshift } &
  \colhead{ Class } &
  \colhead{ log $\nu_p$ } &
  \colhead{ log$\rm L_p$  } &
  \colhead{ log$\rm L_b$  }&
  \colhead{ log$\rm L_R$  }&
  \colhead{ log$\rm L_O$  }&
  \colhead{ log$\rm L_X$  }&
  \colhead{ log$\rm L_{\gamma}$  }&
   \colhead{ $\delta_R$  }&
   \colhead{ Ref  }
    }
 \startdata

3FGL J0050.6-0929	&	PKS 0048-09	&	0.635 	&	IBL	&	14.60 	&	45.99 	&	46.46 	&	43.05 	&	45.75 	&	45.30 	&	46.04 	&	9.6	&	 H09	\\
3FGL J0112.1+2245	&	S2 0109+22	&	0.265 	&	IBL	&	14.39 	&	45.40 	&	45.73 	&	41.94 	&	45.36 	&	44.53 	&	45.41 	&	9.1	&	 S10	\\
3FGL J0222.6+4301	&	3C 66A	&	0.444 	&	IBL	&	14.76 	&	45.94 	&	46.39 	&	43.18 	&	46.21 	&	44.98 	&	46.26 	&	2.6	&	 H09	\\
3FGL J0238.6+1636	&	AO 0235+164	&	0.940 	&	LBL	&	13.24 	&	46.55 	&	46.82 	&	43.78 	&	46.08 	&	45.15 	&	46.89 	&	23.8	 &	S10	\\
3FGL J0303.6+4716	&	4C +47.08	&	0.475 	&	IBL	&	14.10 	&	45.77 	&	46.17 	&	42.86 	&	45.83 	&	43.87 	&	45.62 	&	4.33	 &	F09	\\
3FGL J0424.7+0035	&	PKS 0422+00	&	0.310 	&	IBL	&	14.22 	&	45.57 	&	45.94 	&	42.19 	&	45.16 	&	44.15 	&	45.04 	&	6.11	 &	F09	\\
3FGL J0721.9+7120	&	S5 0716+71	&	0.300 	&	IBL	&	14.96 	&	46.00 	&	46.39 	&	42.32 	&	45.86 	&	44.64 	&	45.95 	&	10.8	 &	S10	\\
3FGL J0738.1+1741	&	PKS 0735+17	&	0.424 	&	IBL	&	14.23 	&	46.07 	&	46.45 	&	43.13 	&	45.69 	&	44.53 	&	45.74 	&	3.92	 &	F09	\\
3FGL J0757.0+0956	&	PKS 0754+100	&	0.266 	&	IBL	&	14.05 	&	45.51 	&	45.86 	&	42.34 	&	45.01 	&	44.14 	&	44.88 	&	 5.5	&	S10	\\
3FGL J0811.3+0146	&	OJ 014	&	1.148 	&	LBL	&	13.28 	&	45.94 	&	46.31 	&	43.44 	&	46.00 	&		&	46.56 	&	5.39	&	 F09	\\
3FGL J0818.2+4223	&	S4 0814+42	&	0.530 	&	IBL	&	13.52 	&	45.26 	&	45.72 	&	43.02 	&	44.95 	&	44.39 	&	46.03 	&	4.6	&	 S10	\\
3FGL J0820.9-1258	&	PKS 0818-128	&	0.074 	&	IBL	&	14.77 	&	43.33 	&	43.93 	&	41.25 	&	43.83 	&	42.83 	&	42.84 	&	 3.18	&	F09	\\
3FGL J0831.9+0430	&	PKS 0829+046	&	0.174 	&	IBL	&	13.84 	&	45.10 	&	45.41 	&	42.06 	&	45.33 	&	43.17 	&	44.68 	&	 3.8	&	F09	\\
3FGL J0854.8+2006	&	OJ 287	&	0.306 	&	IBL	&	14.21 	&	45.87 	&	46.25 	&	42.66 	&	45.45 	&	44.04 	&	45.49 	&	16.8	&	 S10	\\
3FGL J0958.6+6534	&	S4 0954+65	&	0.368 	&	IBL	&	14.02 	&	45.38 	&	45.77 	&	42.51 	&	45.57 	&	44.22 	&	45.18 	&	5.93	 &	F09	\\
3FGL J1058.5+0133	&	4C +01.28	&	0.890 	&	IBL	&	13.79 	&	46.24 	&	46.73 	&	43.95 	&	45.96 	&	45.10 	&	46.71 	&	12.1	 &	S10	\\
3FGL J1221.4+2814	&	W Comae	&	0.103 	&	IBL	&	14.83 	&	44.70 	&	45.09 	&	41.37 	&	44.93 	&	43.18 	&	44.27 	&	1.2	&	 H09	\\
3FGL J1309.5+1154	&	4C +12.46	&	0.415 	&	IBL	&	13.72 	&	44.74 	&	45.26 	&	42.69 	&	44.82 	&	43.94 	&	44.82 	&	1.22	 &	F09	\\
3FGL J1419.9+5425	&	OQ 530	&	0.153 	&	IBL	&	14.27 	&	44.87 	&	45.25 	&	41.75 	&	45.21 	&	43.40 	&	44.10 	&	2.79	&	 F09	\\
3FGL J1540.8+1449	&	4C +14.60	&	0.605 	&	IBL	&	13.97 	&	45.32 	&	45.88 	&	43.24 	&	45.52 	&	44.93 	&	45.13 	&	4.3	&	 S10	\\
3FGL J1719.2+1744	&	PKS 1717+177	&	0.137 	&	IBL	&	13.91 	&	44.02 	&	44.45 	&	41.49 	&	43.93 	&	42.91 	&	44.09 	&	 1.94	&	F09	\\
3FGL J1748.6+7005	&	S4 1749+70	&	0.770 	&	IBL	&	14.27 	&	45.93 	&	46.39 	&	43.18 	&	46.13 	&	45.04 	&	46.28 	&	3.75	 &	F09	\\
3FGL J1751.5+0939	&	OT 081	&	0.322 	&	LBL	&	12.99 	&	45.50 	&	45.76 	&	42.32 	&	45.30 	&	44.07 	&	45.40 	&	11.9	&	 S10	\\
3FGL J1800.5+7827	&	S5 1803+784	&	0.680 	&	IBL	&	13.90 	&	46.18 	&	46.59 	&	43.55 	&	46.19 	&	45.01 	&	46.33 	&	12.1	 &	S10	\\
3FGL J1806.7+6949	&	3C 371	&	0.051 	&	IBL	&	14.60 	&	44.05 	&	44.51 	&	41.16 	&	45.40 	&	42.99 	&	43.58 	&	1.1	&	 S10	\\
3FGL J1824.2+5649	&	4C +56.27	&	0.664 	&	LBL	&	13.25 	&	45.64 	&	46.04 	&	43.33 	&	45.75 	&	45.12 	&	46.03 	&	6.3	&	 S10	\\
3FGL J2005.2+7752	&	S5 2007+77	&	0.342 	&	IBL	&	13.55 	&	45.08 	&	45.49 	&	42.61 	&	44.60 	&	44.18 	&	45.14 	&	4.68	 &	F09	\\
3FGL J2134.1-0152	&	PKS 2131-021	&	1.283 	&	LBL	&	13.17 	&	46.10 	&	46.55 	&	43.99 	&	45.65 	&	45.30 	&	46.38 	&	7	 &	F09	\\
3FGL J2202.7+4217	&	BL Lacertae	&	0.069 	&	IBL	&	15.10 	&	44.78 	&	45.24 	&	41.92 	&	44.80 	&	43.44 	&	44.52 	&	7.2	&	 S10	\\
3FGL J2236.3+2829	&	B2 2234+28A	&	0.795 	&	LBL	&	12.88 	&	45.73 	&	46.02 	&	43.39 	&	45.52 	&		&	46.37 	&	6	&	 H09	\\
3FGL J0108.7+0134	&	4C +01.02	&	2.099 	&	IF	&	13.53 	&	46.47 	&	47.00 	&	44.57 	&	46.29 	&	45.47 	&	47.66 	&	18.2	 &	S10	\\
3FGL J0137.0+4752	&	OC 457	&	0.859 	&	LF	&	12.69 	&	46.14 	&	46.41 	&	43.47 	&	45.50 	&	45.05 	&	46.55 	&	20.5	&	 S10	\\
3FGL J0151.6+2205	&	PKS 0149+21	&	1.320 	&	LF	&	13.14 	&	46.09 	&	46.48 	&	43.80 	&	45.59 	&		&	46.23 	&	4.72	&	 LV99	\\
3FGL J0217.5+7349	&	S5 0212+73	&	2.367 	&	LF	&	13.35 	&	46.95 	&	47.31 	&	44.60 	&	46.15 	&	46.23 	&	47.47 	&	8.4	&	 S10	\\
3FGL J0217.8+0143	&	PKS 0215+015	&	1.715 	&	IF	&	14.66 	&	46.83 	&	47.27 	&	43.87 	&	45.93 	&	45.72 	&	47.26 	&	 5.61	&	F09	\\
3FGL J0237.9+2848	&	4C +28.07	&	1.213 	&	IF	&	13.59 	&	46.75 	&	47.10 	&	44.05 	&	46.17 	&	45.47 	&	47.23 	&	16	&	 S10	\\
3FGL J0309.0+1029	&	PKS 0306+102	&	0.863 	&	IF	&	14.04 	&	46.68 	&	46.96 	&	43.12 	&	44.36 	&	45.21 	&	46.25 	&	 2.79	&	F09	\\
3FGL J0336.5+3210	&	NRAO 140	&	1.259 	&	IF	&	13.55 	&	46.28 	&	46.72 	&	44.17 	&	46.63 	&	46.43 	&	46.62 	&	22	&	 S10	\\
3FGL J0339.5-0146	&	PKS 0336-01	&	0.850 	&	LF	&	13.40 	&	46.12 	&	46.51 	&	43.79 	&	45.89 	&	44.48 	&	46.42 	&	17.2	 &	S10	\\
3FGL J0423.2-0119	&	PKS 0420-01	&	0.916 	&	LF	&	12.88 	&	46.60 	&	46.84 	&	43.90 	&	46.51 	&	45.69 	&	46.70 	&	19.7	 &	S10	\\
3FGL J0449.0+1121	&	PKS 0446+11	&	1.207 	&	LF	&	13.09 	&	45.98 	&	46.35 	&	43.63 	&	45.39 	&	45.06 	&	46.72 	&	4.9	&	 LV99	\\
3FGL J0501.2-0157	&	S3 0458-02	&	2.286 	&	LF	&	13.50 	&	46.57 	&	47.03 	&	44.58 	&	46.33 	&	45.99 	&	47.53 	&	15.7	 &	S10	\\
3FGL J0530.8+1330	&	PKS 0528+134	&	2.070 	&	LF	&	12.53 	&	46.78 	&	47.03 	&	44.34 	&	45.91 	&	46.45 	&	47.47 	&	 30.9	&	S10	\\
3FGL J0608.0-0835	&	PKS 0605-08	&	0.872 	&	IF	&	13.88 	&	46.13 	&	46.58 	&	43.70 	&	46.31 	&	45.09 	&	46.32 	&	7.5	&	 S10	\\
3FGL J0739.4+0137	&	PKS 0736+01	&	0.189 	&	IF	&	14.43 	&	45.04 	&	45.51 	&	42.34 	&	44.78 	&	44.32 	&	44.73 	&	8.5	&	 S10	\\
3FGL J0807.9+4946	&	OJ 508	&	1.434 	&	LF	&	13.28 	&	46.12 	&	46.50 	&	43.90 	&	47.49 	&	45.39 	&	46.15 	&	35.2	&	 S10	\\
3FGL J0830.7+2408	&	OJ 248	&	0.939 	&	LF	&	13.50 	&	46.26 	&	46.59 	&	43.36 	&	46.04 	&	45.60 	&	46.27 	&	13	&	 S10	\\
3FGL J0841.4+7053	&	S5 0836+71	&	2.218 	&	IF	&	14.44 	&	47.01 	&	47.53 	&	44.78 	&	47.16 	&	46.91 	&	47.39 	&	16.1	 &	S10	\\
3FGL J0850.2-1214	&	PMN J0850-1213	&	0.566 	&	LF	&	13.10 	&	45.84 	&	46.02 	&	42.55 	&	45.40 	&		&	45.89 	&	16.5	 &	H09	\\
3FGL J0948.6+4041	&	4C +40.24	&	1.249 	&	IF	&	13.86 	&	46.25 	&	46.72 	&	43.94 	&	45.90 	&	44.90 	&	46.08 	&	6.3	&	 S10	\\
3FGL J0956.6+2515	&	OK 290	&	0.708 	&	IF	&	13.98 	&	45.93 	&	46.34 	&	43.27 	&	45.29 	&	44.77 	&	45.74 	&	4.3	&	 H09	\\
3FGL J0957.6+5523	&	4C +55.17	&	0.899 	&	IF	&	14.74 	&	45.76 	&	46.40 	&	43.94 	&	45.91 	&	44.97 	&	46.73 	&	4.63	 &	LV99	\\
3FGL J1037.0-2934	&	PKS 1034-293	&	0.312 	&	IF	&	13.92 	&	45.39 	&	45.78 	&	42.55 	&	44.71 	&	44.25 	&	44.61 	&	 2.8	&	F09	\\
3FGL J1129.9-1446	&	PKS 1127-14	&	1.184 	&	IF	&	13.99 	&	46.42 	&	46.93 	&	44.44 	&	46.65 	&	45.73 	&	46.56 	&	3.22	 &	F09	\\
3FGL J1159.5+2914	&	Ton 599	&	0.725 	&	LF	&	13.04 	&	45.89 	&	46.29 	&	43.57 	&	45.68 	&	45.03 	&	46.54 	&	28.2	&	 S10	\\
3FGL J1224.9+2122	&	4C +21.35	&	0.432 	&	IF	&	14.53 	&	45.68 	&	46.16 	&	43.12 	&	45.60 	&	44.92 	&	46.52 	&	5.2	&	 S10	\\
3FGL J1229.1+0202	&	3C 273	&	0.158 	&	IF	&	15.12 	&	45.92 	&	46.52 	&	43.63 	&	45.53 	&	45.38 	&	45.20 	&	16.8	&	 S10	\\
3FGL J1256.1-0547	&	3C 279	&	0.536 	&	LF	&	12.69 	&	46.39 	&	46.76 	&	43.98 	&	45.71 	&	46.15 	&	46.68 	&	23.8	&	 S10	\\
3FGL J1310.6+3222	&	OP 313	&	0.998 	&	LF	&	13.22 	&	46.72 	&	47.03 	&	43.77 	&	45.39 	&	45.12 	&	46.57 	&	15.3	&	 S10	\\
3FGL J1326.8+2211	&	B2 1324+22	&	1.400 	&	LF	&	12.97 	&	46.35 	&	46.70 	&	43.76 	&	45.94 	&	45.29 	&	46.74 	&	21	&	 S10	\\
3FGL J1337.6-1257	&	PKS 1335-127	&	0.539 	&	LF	&	13.25 	&	46.34 	&	46.68 	&	43.42 	&	45.69 	&	45.18 	&	45.64 	&	 8.3	&	S10	\\
3FGL J1408.8-0751	&	PKS B1406-076	&	1.494 	&	LF	&	12.86 	&	46.11 	&	46.43 	&	43.72 	&	45.72 	&	45.05 	&	46.88 	&	 8.26	&	LV99	\\
3FGL J1504.4+1029	&	PKS 1502+106	&	1.838 	&	LF	&	13.34 	&	46.62 	&	47.00 	&	44.30 	&	46.03 	&	45.03 	&	48.01 	&	 11.9	&	S10	\\
3FGL J1512.8-0906	&	PKS 1510-08	&	0.360 	&	IF	&	13.97 	&	45.47 	&	45.94 	&	43.06 	&	45.40 	&	44.18 	&	46.59 	&	16.5	 &	S10	\\
3FGL J1608.6+1029	&	4C +10.45	&	1.226 	&	LF	&	13.39 	&	46.15 	&	46.59 	&	43.86 	&	45.87 	&	45.46 	&	46.66 	&	24.8	 &	S10	\\
3FGL J1613.8+3410	&	OS 319	&	1.399 	&	LF	&	13.44 	&	46.56 	&	46.98 	&	44.43 	&	46.51 	&	45.38 	&	46.35 	&	13.6	&	 S10	\\
3FGL J1635.2+3809	&	4C +38.41	&	1.813 	&	LF	&	13.21 	&	46.88 	&	47.24 	&	44.48 	&	46.63 	&	45.25 	&	47.94 	&	21.3	 &	S10	\\
3FGL J1637.9+5719	&	OS 562	&	0.751 	&	IF	&	14.22 	&	46.07 	&	46.53 	&	43.37 	&	45.85 	&	45.16 	&	45.50 	&	13.9	&	 S10	\\
3FGL J1642.9+3950	&	3C 345	&	0.593 	&	LF	&	13.46 	&	46.15 	&	46.61 	&	43.93 	&	45.64 	&	45.26 	&	45.99 	&	7.7	&	 S10	\\
3FGL J1728.5+0428	&	PKS 1725+044	&	0.296 	&	LF	&	13.32 	&	44.94 	&	45.25 	&	42.29 	&	45.10 	&	43.83 	&	44.84 	&	 3.8	&	H09	\\
3FGL J1733.0-1305	&	PKS 1730-13	&	0.902 	&	LF	&	12.62 	&	46.12 	&	46.51 	&	44.23 	&	44.93 	&	45.86 	&	46.70 	&	10.6	 &	S10	\\
3FGL J1740.3+5211	&	4C +51.37	&	1.375 	&	LF	&	13.42 	&	46.29 	&	46.68 	&	43.72 	&	46.50 	&	45.64 	&	46.80 	&	26.3	 &	S10	\\
3FGL J1744.3-0353	&	PKS 1741-03	&	1.054 	&	IF	&	14.06 	&	46.89 	&	47.27 	&	43.74 	&	45.59 	&	45.84 	&	45.95 	&	19.5	 &	S10	\\
3FGL J1924.8-2914	&	PKS B1921-293	&	0.352 	&	LF	&	12.53 	&	45.88 	&	46.14 	&	43.74 	&	45.33 	&	44.94 	&	45.38 	&	 9.51	&	F09	\\
3FGL J2123.6+0533	&	OX 036	&	1.941 	&	LF	&	13.40 	&	46.91 	&	47.25 	&	43.99 	&	45.73 	&	45.80 	&	46.46 	&	15.2	&	 S10	\\
3FGL J2147.2+0929	&	PKS 2144+092	&	1.113 	&	IF	&	13.87 	&	46.46 	&	46.81 	&	43.62 	&	46.21 	&	45.04 	&	46.67 	&	 5.96	&	LV99	\\
3FGL J2158.0-1501	&	PKS 2155-152	&	0.672 	&	LF	&	13.09 	&	45.60 	&	46.00 	&	43.67 	&	45.41 	&	44.99 	&	45.67 	&	 2.31	&	F09	\\
3FGL J2203.7+3143	&	4C +31.63	&	0.295 	&	IF	&	14.43 	&	45.48 	&	45.96 	&	42.91 	&	45.79 	&	44.96 	&	44.37 	&	6.6	&	 S10	\\
3FGL J2225.8-0454	&	3C 446	&	1.404 	&	LF	&	13.24 	&	46.77 	&	47.20 	&	44.70 	&	46.52 	&	45.83 	&	46.83 	&	15.9	&	 S10	\\
3FGL J2229.7-0833	&	PKS 2227-08	&	1.560 	&	LF	&	13.34 	&	46.76 	&	47.09 	&	43.91 	&	46.23 	&	46.16 	&	47.24 	&	15.8	 &	S10	\\
3FGL J2232.5+1143	&	CTA 102	&	1.037 	&	IF	&	13.65 	&	46.41 	&	46.88 	&	44.43 	&	46.30 	&	45.68 	&	46.88 	&	15.5	&	 S10	\\
3FGL J2254.0+1608	&	3C 454.3	&	0.859 	&	IF	&	13.54 	&	46.71 	&	47.16 	&	44.51 	&	46.56 	&	46.30 	&	47.49 	&	32.9	 &	S10	\\
3FGL J0205.0+1510	&	4C +15.05	&	0.405 	&	LBCU	&	12.10 	&	44.53 	&	44.99 	&	43.35 	&	43.50 	&	42.53 	&	45.03 	&	15	 &	S10	\\
3FGL J0522.9-3628	&	PKS 0521-36	&	0.057 	&	IBCU	&	13.75 	&	44.23 	&	44.60 	&	42.16 	&	45.46 	&	43.69 	&	43.87 	&	 1.83	&	F09	\\
3FGL J0725.8-0054	&	PKS 0723-008	&	0.128 	&	IBCU	&	14.00 	&	44.34 	&	44.81 	&	41.84 	&	44.31 	&	43.12 	&	43.94 	&	 2.5	&	LV99	\\
3FGL J1416.0+1325	&	PKS B1413+135	&	0.247 	&	LBCU	&	12.57 	&	44.73 	&	44.97 	&	42.32 	&	43.70 	&	42.33 	&	44.34 	&	 12.1	&	S10	\\
\enddata

Note to the Table:
Col. (1) gives the 3FGL name;
Col. (2) Other name;
Col. (3) redshift from NED database at IPAC;
Col. (4)  the SED  classification  by our method (Fan et al. 2016a)
  F stands for  FSRQ,
  B for BL Lac;
Col. (5)   peak frequency, $log \nu_p$ (Hz);
Col. (6)  peak luminosity, $\rm log L_p$ (erg/s);
Col. (7)  bolometric luminosity, $\rm log L_{bol}$ (erg/s);
Col. (8)  radio luminosity, $\rm log L_R$ (erg/s);
Col. (9)  optical luminosity, $\rm log L_O$ (erg/s);
Col. (10)  X-ray luminosity, $\rm log L_X$ (erg/s);
 Col. (11)  $\gamma$-ray luminosity, $\rm log L_{\gamma}$ (erg/s);
Col. (12)  Doppler factor, $\rm \delta_R$; and
Col. (13)  reference for Doppler factor.
          F09: Fan et al. (2009);
          H09: Hovatta et al. (2009);
          LV99: L\"ahteenim\"aki \& Valtaoja (1999);
           S10: Savolainen et al. (2010)
\label{sample}
\end{deluxetable}

\end{document}